# Low Resistance Ohmic Contact to P-type Monolayer WSe$_2$

Jingxu Xie[1,2,3,*], Zuocheng Zhang[1,*], Haodong Zhang[1,*], Vikram Nagarajan[1], Wenyu Zhao[1], Haleem Kim[1,3], Collin Sanborn[1], Ruishi Qi[1,3], Sudi Chen[1], Salman Kahn[1], Kenji Watanabe[4], Takashi Taniguchi[5], Alex Zettl[1,3,6], Michael Crommie[1,3,6], James Analytis[1,3,6], Feng Wang[1,3,6]†

[1] Department of Physics, University of California at Berkeley, Berkeley, California 94720, United States

[2] Graduate Group in Applied Science and Technology, University of California at Berkeley, Berkeley, California 94720, United States

[3] Material Science Division, Lawrence Berkeley National Laboratory, Berkeley, California 94720, United States

[4] Research Center for Electronic and Optical Materials, National Institute for Materials Science, 1-1 Namiki, Tsukuba 305-0044, Japan

[5] Research Center for Materials Nanoarchitectonics, National Institute for Materials Science, 1-1 Namiki, Tsukuba 305-0044, Japan

[6] Kavli Energy NanoSciences Institute at University of California Berkeley and Lawrence Berkeley National Laboratory, Berkeley, California 94720, United States

* These authors contributed equally to this work

† Correspondence to: fengwang76@berkeley.edu

**Abstract**

Advanced microelectronics in the future may require semiconducting channel materials beyond silicon. Two-dimensional (2D) semiconductors, characterized by their atomically thin thickness, hold immense promise for high-performance electronic devices at the nanometer scale with lower heat dissipation. One challenge for achieving high-performance 2D semiconductor field effect transistors (FET), especially for p-type materials, is the high electrical contact resistance present at the metal-semiconductor interface. In conventional bulk semiconductors, low resistance ohmic contact is realized through heavy substitutional doping with acceptor or donor impurities at the contact region. The strategy of substitutional doping, however, does not work for p-type 2D semiconductors such as monolayer tungsten diselenide ($WSe_2$). In this study, we developed highly efficient charge-transfer doping with $WSe_2$/α-$RuCl_3$ heterostructures to achieve low-resistance ohmic contact for p-type $WSe_2$ transistors. We show that a hole doping as high as $3 \times 10^{13}$ cm$^{-2}$ can be achieved in the $WSe_2$/α-$RuCl_3$ heterostructure due to its type-III band alignment. It results in an Ohmic contact with resistance lower than 4 kΩ μm at the p-type monolayer $WSe_2$/metal junction. Using this low-resistance contact, we demonstrate high-performance p-type $WSe_2$ transistors with a saturation current of 35 μA·μm$^{-1}$ and an $I_{ON}/I_{OFF}$ ratio exceeding $10^9$ at room temperature. It could enable future microelectronic devices based on 2D semiconductors and contribute to the extension of Moore's law.

**Main Text**

The family of transition-metal dichalcogenide (TMD) materials possesses exceptional characteristics, including atomically thin structures and the absence of dangling bonds, that make them ideal candidates for advanced electronic applications[1-12]. However, it is challenging to form low-resistance electrical contact to TMD monolayers due to the Schottky barrier between the three-dimensional metal and the two-dimensional (2D) semiconductors. This issue can limit the ultimate scaling and performance of TMD-based 2D electronic devices[3, 13-17]. Heavy substitutional doping, widely used to achieve low resistance ohmic contact to bulk semiconductors, does not work well for atomically thin 2D materials[3]. A variety of new contact strategies have been explored to realize ohmic contact to TMD materials[18-20]. For n-type TMD monolayers such as $MoS_2$, there has been impressive progress where a contact resistance as low as 123 Ω μm has been demonstrated[20]. In contrast, low-resistance Ohmic contact to p-type monolayer semiconductors remains a challenge. The lowest reported contact resistance to p-type monolayer $WSe_2$, for example, is limited to 229 kΩ μm[18]. For complementary metal-oxide semiconductor (CMOS) devices in advanced microelectronics, both n-type and p-type field effect transistors with low contact resistances are necessary.

Here we present a new approach to achieving Ohmic contact to monolayer $WSe_2$ and demonstrate high performance p-type $WSe_2$ field-effect transistor. Our approach utilizes efficient charge-transfer doping in the $WSe_2$/α-$RuCl_3$ heterostructure with type-III band alignment. Figure 1a illustrates the type-III band alignment between monolayer $WSe_2$ and α-$RuCl_3$, wherein the conduction band minimum of α-$RuCl_3$ is positioned lower than the valence band maximum of monolayer WSe$_2$. The energy difference between these two levels is approximately 0.8 eV[21]. When WSe$_2$ is in close proximity to α-RuCl$_3$, a charge transfer phenomenon occurs. As a result, the WSe$_2$ contact region becomes heavily hole-doped and highly conductive. Previously, highly efficient charge transfer doping of graphene by α-RuCl$_3$ has been observed[22-24]. Recent optical spectroscopy shows that strong charge transfer doping of WSe$_2$ is present in WSe$_2$/α-RuCl$_3$ heterostructures[25] and this doping could improve electrical contact to the WSe$_2$ layer[26]. To examine the contact resistance of monolayer WSe$_2$ with an α-RuCl$_3$ charge transfer interface, we fabricated a monolayer WSe$_2$ device for the transmission line method[27] (TLM) measurement, which is a commonly used technique for determining the contact resistance in electronic devices. Figure 1b shows the schematic side view of the TLM device and Figure 1c presents the optical image of the device where the sample width is 7 μm (WSe$_2$ is outlined by the blue solid line). The 10nm-thick platinum electrodes[28-30] were pre-patterned on a SiO$_2$/Si substrate with separation ranging from 0.3 μm to 3 μm. A monolayer WSe$_2$ that was fully covered by a few-layer α-RuCl$_3$ flake (outlined by the orange solid line) was then released on top of the platinum electrodes.

The I-V curves of different electrodes at room and low temperatures were measured and plotted in Figure 2a, b. Well-defined linear characteristics were observed at both room temperature (Figure 2a) and low temperature (Figure 2b), indicating the absence of the Schottky barrier at the contact interface and confirming the presence of Ohmic contacts. The contact resistance values extracted from the TLM method at different temperatures are presented in Figure 2c, d, showing the contact resistance values ($R_C$) of 4 kΩ μm at room temperature and 4.5 kΩ μm at low temperature (extracted after subtracting the resistance of Platinum electrodes), suggesting the robustness and consistency of the contact interface. This is a very low p-type contact resistance for monolayer $WSe_2$ devices. We also evaluated the contact resistance using gold/graphite/$WSe_2$/α-$RuCl_3$ contacts (Extended Data Figure 1) and similar results were observed. Such a low contact resistance value is desirable for efficient charge carrier injection, therefore enabling better device performance and overall functionality. We also note that the α-$RuCl_3$ flake we used itself has a resistance of about 100 kΩ μm at room temperature and >1 MΩ μm at low temperature[31], thus its contribution to the contact resistance is negligible.

To quantify the charge transfer doping level in the $WSe_2$/α-$RuCl_3$ heterostructure, the Hall measurement was employed. We fabricated monolayer $WSe_2$/α-$RuCl_3$ devices with standard Hall-bar electrodes (Figure 3a) which are made of prepatterned few-layer graphite (outlined by the purple lines). Figure 3b shows the Hall resistance $R_{xy}$ as a function of perpendicular magnetic field $B$ at room and low temperatures. The $R_{xy}$ increases linearly with the perpendicular magnetic field. The positive sign of the Hall slope confirms the hole doping in the monolayer $WSe_2$. From the linear fit to Hall resistance, we estimate a hole carrier density of $3.1 \times 10^{13}$ $cm^{-2}$ at room temperature and $3.3 \times 10^{13}$ $cm^{-2}$ at low temperature. The same behavior and similar charge-transfer hole density are observed in a second monolayer $WSe_2$/α-$RuCl_3$ Hall bar device (Extended Data Figure 2). This confirms that the monolayer $WSe_2$ contact region has undergone heavy hole doping through the charge transfer mechanism with α-$RuCl_3$. This doping level is significantly higher than what can typically be achieved by conventional gate injection methods using hBN as the dielectric[32], where the hole carrier density is typically less than $1\times10^{13}$ cm$^{-2}$.

The low contact resistance to monolayer WSe$_2$ enables us to achieve high performance p-type monolayer WSe$_2$ field-effect transistors (FETs). Figure 4a depicts the schematic side view of an FET device and Figure 4b presents the corresponding optical image. The contact regions of WSe$_2$ are positioned between few-layer graphite and few-layer α-RuCl$_3$, forming a sandwich structure. The α-RuCl$_3$ flakes were pre-patterned by AFM cutting[33] to make an FET channel length of around 0.5 μm. Then chromium and gold (typically 5 nm and 50nm) were sequentially deposited on the few-layer graphite contacts to make the electrodes.

The switching behavior of our monolayer WSe$_2$ FET was characterized at room temperature in the configuration presented in Figure 4a. A few-layer graphene and a 30 nm hBN were used as a top gate and a gate dielectric, respectively. We swept both the top gate voltage ($V_{GS}$) and the source-drain bias voltage ($V_{DS}$) across the monolayer WSe$_2$ conductive channel. Figure 4c shows the 2D color plot of drain-source current $I_{DS}$ as a function of $V_{DS}$ and $V_{GS}$ at room temperature. A typical FET behavior is observed: as the top gate voltages $V_{GS}$ increases, the FET undergoes a transition from the off state to the on state. The on-state drain-source current can be as high as 35 μA μm$^{-1}$ at $V_{GS}$=16 V, where the gate-induced carrier density in the monolayer WSe$_2$ channel is $0.7\times10^{13}$ cm$^{-2}$. The overall two-terminal resistance is 28 kΩ μm at room temperature.

Figure 4d depicts the measured drain-source $I_{DS}$-$V_{DS}$ curves, corresponding to the horizontal line cuts in Figure 4c. The source-drain current $I_{DS}$ varies linearly with $V_{DS}$ at sufficiently high gate voltages. The reliable and Ohmic nature of the contact is crucial for the practical utilization of the device in electronic applications, providing confidence in its performance and potential for use in various technological applications.

Figure 4e presents the transfer characteristics of the monolayer $WSe_2$ FET device, corresponding to the vertical line cuts in Figure 4c. The observed current variation for different values of top gate voltages $V_{GS}$ suggests that the field-effect behavior of our transistor is primarily governed by the monolayer $WSe_2$ channel rather than the contacts. For a drain-source bias voltage $V_{DS}$ = 100mV, we observe an on-current of 4 µA µm$^{-1}$. At drain-source bias voltage $V_{DS}$ = 1V, the maximal measured on-current is 35 µA µm$^{-1}$, indicating the device's ability to efficiently conduct current in the on state. Additionally, the device exhibits an ultralow off-state current of approximately $10^{-8}$ µA µm$^{-1}$, making it ideal for applications in devices with ultralow standby power dissipation. The measured drain current modulation, with an on/off current ratio $I_{ON}/I_{OFF}$ exceeding $10^9$, highlights the exceptional performance of the monolayer $WSe_2$ transistor with α-$RuCl_3$ contacts. Such a high on/off current ratio implies that the device can achieve rapid switching and short latency due to the high on-state current. Simultaneously, the low off-state current ensures minimal static power consumption, making it highly promising for applications in digital electronics.

The subthreshold swing is a measure of the efficiency of a transistor in controlling the current flow when it operates in the subthreshold region[34]. We have observed a subthreshold swing of approximately 0.7 V per decade in our monolayer WSe$_2$ transistor with α-RuCl3 contacts, which is higher compared to that in commercial silicon-based devices (~70 mV per decade at room temperature). The rather big subthreshold swing is mainly attributed to the large thickness of the hBN dielectric (~30 nm) used in this device. Further optimization of using a thinner hBN of about 15nm can reduce the threshold swing to approximately 0.2 V per decade (Extended Data Figure 3). This improvement suggests that the choice of gate dielectric can play a crucial role in optimizing the subthreshold swing and overall device performance. Therefore we can further reduce the subthreshold swing by using a gate dielectric with thinner thickness or higher dielectric constant, such as $HfO_2$[35] or $Bi_2SeO_5$[36], allowing for more efficient modulation of the carrier density in the monolayer WSe$_2$ channel.

For potential applications in digital and radiofrequency devices, saturation of the drain current is crucial for achieving maximum operation speeds[37]. At room temperature, a well-defined current saturation can be achieved at all gate voltages within the high drain-source bias region (Figure 4f). And the saturation current can be as high as 39 µA µm$^{-1}$. Furthermore, the electrical contacts exhibit Ohmic behavior within the linear region at low drain-source biases. The presence of such a well-developed saturation behavior, which is not typically observed in graphene-based FET devices[37], is important for achieving high power gains. Because our channel material is a 0.7 nm thick monolayer WSe$_2$[38], it will be resilient against short-channel effects when the channel length is scale down to the nanometer range. Therefore, monolayer

WSe$_2$ with α-RuCl$_3$ contacts could be promising for high-speed field-effect device applications. We note that our device exhibits relatively low on-state conductance and relatively high threshold source-drain bias compared to typical silicon-based devices. These characteristics are mostly limited by the long channel length in our current device. By reducing the channel length to the nanometer scale and utilizing a thinner gate dielectric or a high-k dielectric, we can anticipate improved device performance, including larger saturation current and lower threshold bias. Further investigations are required to explore and evaluate the limits of device performance for monolayer WSe$_2$ FETs with α-RuCl$_3$ contacts.

Figure 5a,b benchmarks the performance of our device against other reported values using various fabrication methods. Here we restrict the comparison with studies of p-type contacts to monolayer WSe$_2$ at room temperature and $|V_{DS}| = 1$ V. To the best of our knowledge, the contact resistance value is among the lowest reported in the literature so far[18, 19, 38-43]. Figure 5b shows the maximum drain current versus the on/off ratio. The drain current and on/off ratio exceeds that obtained by other methods such as the transferred metal-hBN method[39] and Pd contacts[18]. The monolayer p-type WSe$_2$ FET performance is comparable with the best values achieved for n-type FETs with monolayer MoS$_2$[20], indicating the potential to create monolayer semiconducting TMD-based complementary metal-oxide-semiconductor architecture.

To summarize, our work demonstrates the successful realization of low-resistance p-type Ohmic contact to monolayer WSe$_2$ by utilizing α-RuCl$_3$ as a charge transfer interface, which enables the heavy hole doping of the contact region with the hole density of $\sim 3\times 10^{13}$ cm$^{-2}$. The use of α-RuCl$_3$ as a dopant material and the charge transfer mechanism offers a reliable and effective method to achieve the desired p-type conductivity in monolayer WSe$_2$ devices. The contact resistance to monolayer WSe$_2$ can be as low as 4 kΩ μm. And high-performance p-type FET devices were fabricated using this new contact technique. Our samples exhibit a drain saturation current at on-state up to 35 μA μm$^{-1}$, and an on-off ratio of ~10$^9$ at room temperature. Our results could help to realize the next generation of electronics based on monolayer WSe$_2$ and open a new route to probe the novel electrical transport properties of WSe$_2$ based devices.

## Methods

**Sample preparation and device fabrication**. Two-dimensional flakes (monolayer $WSe_2$, graphite, hBN, α-$RuCl_3$) were prepared by mechanical exfoliation from the bulk crystal on a 90nm $SiO_2$/Si substrate via the Scotch tape method. Monolayer and multilayer flakes were identified with optical microscopy. Polypropylene carbon (PPC) and polyethylene terephthalate glycol (PETG) based dry transfer technology were used to subsequently pick up the flakes. The α-$RuCl_3$ flakes were pre-patterned by atomic force microscopy (AFM) cutting to have a gap of around 0.5 μm before the stacking process to make the monolayer $WSe_2$ FET. Then photolithography and electron-beam lithography were used to pattern the electrodes. Chromium and gold were sequentially evaporated on the few-layer graphene contact and gate to make electrodes. For the TLM device, the chromium/platinum (typically 3nm and 10nm) electrodes were pre-patterned on a $SiO_2$/Si substrate before releasing the sample on them.

**Measurements.** The transport measurements were performed in a Cryomagnetics superconducting magnet system with a variable temperature insert. Samples were in a Helium environment of around 0.3 bar. Transport characteristics were mainly measured by applying DC voltage with the Keithley 2612B SourceMeter and Keithley 6482 picoammeter. The Hall measurements were performed by the standard four-probe AC lock-in method using an SRS 830 lock-in amplifier with an AC current of 10nA and frequency of around 17Hz.

## Data availability

The data that support the findings of this study are available from the corresponding author upon reasonable request.

**Competing interests**

The authors declare that they have no competing interests.

**Acknowledgements**

The electrical transport measurements are supported by the U.S. Department of Energy, Office of Science, Office of Basic Energy Sciences, Materials Sciences and Engineering Division (DE-AC02-05-CH11231), within the van der Waals Heterostructure Program (KCWF16). The lithography patterning is supported by Army Research Office award W911NF2110176. The WSe2 monolayer-RuCl3 heterostructure fabrication is supported by the U.S. Department of Energy, Office of Science, National Quantum Information Science Research Centers, Quantum Systems Accelerator. K.W. and T.T. acknowledge support from the JSPS KAKENHI (Grant Numbers 21H05233 and 23H02052) and World Premier International Research Center Initiative (WPI), MEXT, Japan.

**Author contributions**

F. W. conceived the research. F. W. and J. A. supervised the project. J. X., H. Z., and Z. Z. fabricated the device and performed most of the experimental measurements together. W. Z., C. S., R. Q., S. C., S. K., M. C., and A. Z. contributed to the fabrication of van der Waals heterostructures. H. K. contributed to the electrical transport measurements. J. X., Z. Z., and F. W. performed data analysis. K. W. and T. T. grew hBN crystals. V. N. grew α-RuCl$_3$ crystals. All authors discussed the results and wrote the manuscript.

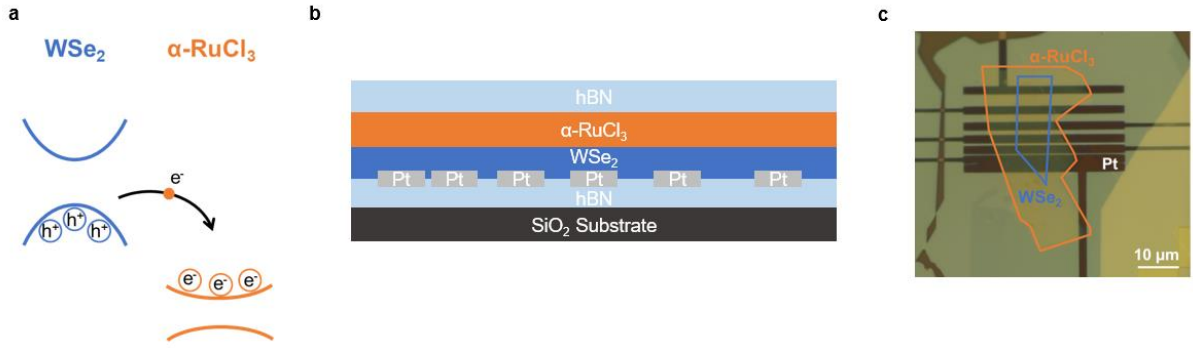

**Figure 1: WSe$_2$/α-RuCl$_3$ band alignment and contact resistance measurements.** (a) Type-III band alignment of WSe$_2$ and α-RuCl$_3$. The conduction band minimum of α-RuCl$_3$ is lower than the valence band maximum of WSe$_2$ and there is a spontaneous charge transfer between WSe$_2$ and α-RuCl$_3$, resulting in a heavily doped WSe$_2$ layer. (**b**) Schematic side view of the device for transmission line method (TLM). (**c**) Optical image of the TLM device. The separation between electrodes ranges from 0.3 μm to 3 μm. The blue and orange shapes mark the monolayer WSe$_2$ and α-RuCl$_3$ region, respectively.

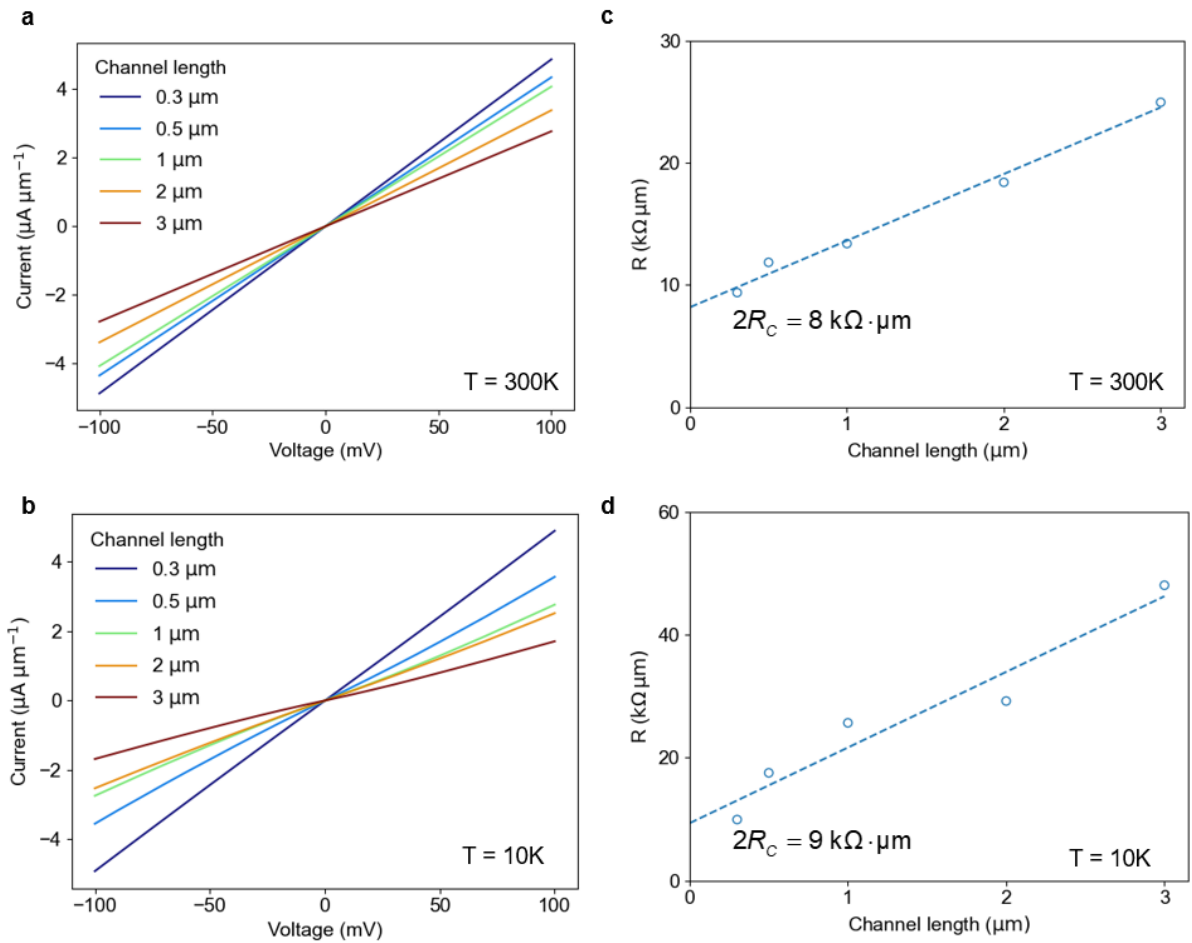

**Figure 2: Contact resistance to a monolayer WSe$_2$ with the Pt/WSe$_2$/α-RuCl$_3$ contact. (a and b)** I-V curves for different channel lengths at room temperature (**a**) and T = 10K (**b**). Linear behavior is observed at both room temperature and low temperature, indicating the absence of a contact barrier. (**c**) and (**d**) Contact resistance $R_C$ extracted using TLM at room temperature (**c**) and T = 10K (**d**).

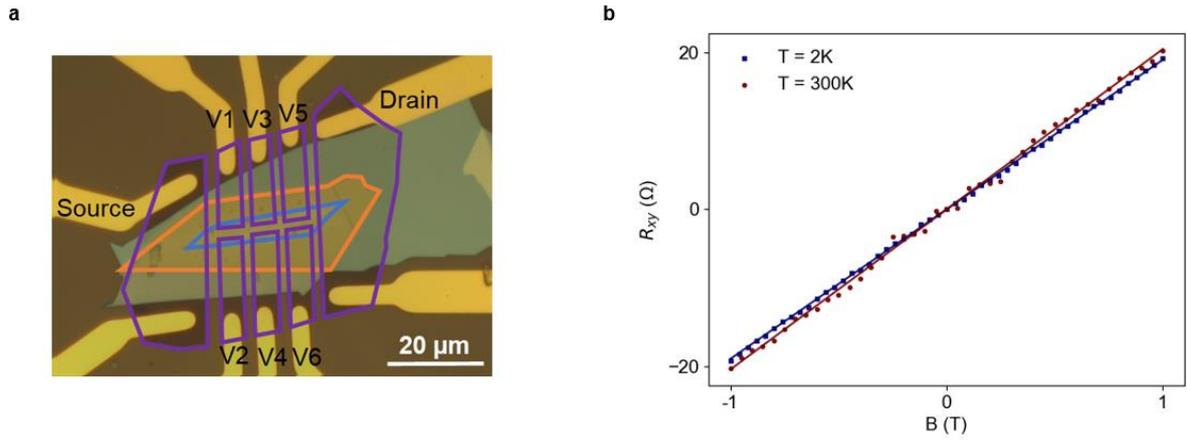

**Figure 3: Determination of carrier density in the charge-transfer doped WSe₂ with α-RuCl₃.** (**a**) Optical image of the multi-terminal monolayer WSe₂/α-RuCl₃ device for the Hall measurements. Blue, orange and purple boxes outline the monolayer WSe₂, α-RuCl₃, and graphite electrodes, respectively. (**b**) Hall resistance $R_{xy}$ as a function of magnetic field where solid lines are linear fits to the experiment data. The positive sign of Hall slope confirms the hole doping in the monolayer WSe₂. We estimate a hole carrier density of $3.1 \times 10^{13}$ cm$^{-2}$ at room temperature and of $3.3 \times 10^{13}$ cm$^{-2}$ at low temperature from the Hall slope.

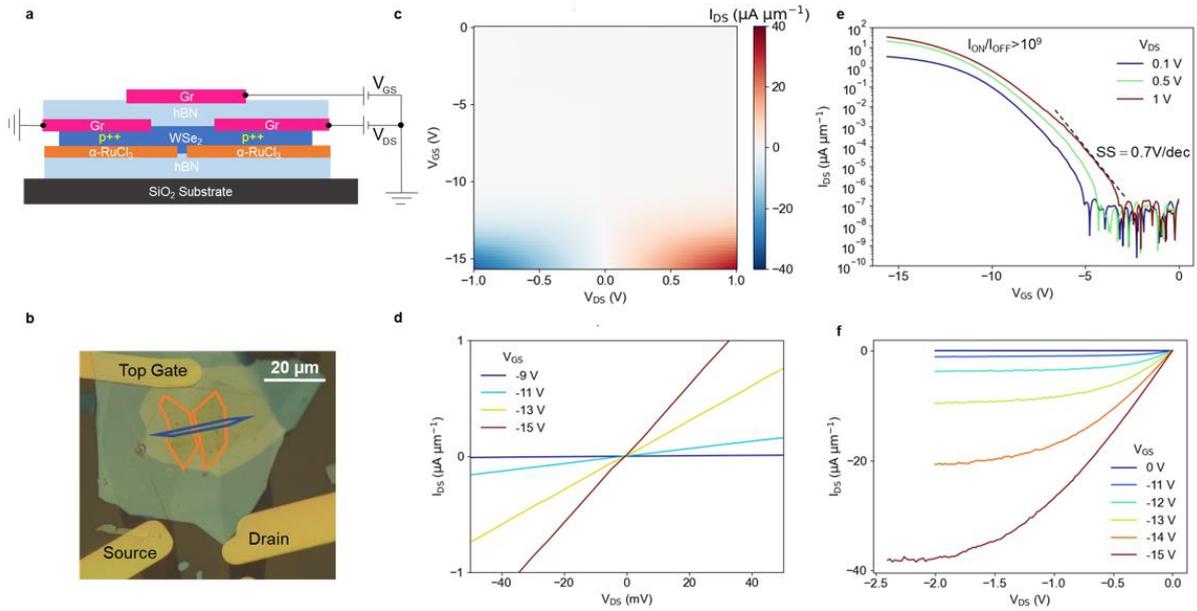

**Figure 4: Characterization of monolayer WSe₂ FET.** (**a**) Schematic cross-sectional view of FET device structure. The contact regions are sandwiched by a few-layer graphite and a few-layer α-RuCl₃. The monolayer WSe₂ channel is separated from the few-layer graphene top gate by a 30-nm-thick hBN (**b**) Optical image of the monolayer WSe₂ FET device. Blue and orange boxes mark the monolayer WSe₂ region and α-RuCl₃ contact region, respectively. The channel length and width are 0.5 µm and 2 µm. (**c**) 2D color plot of drain-source current $I_{DS}$ as a function of drain-source voltage $V_{DS}$ and top gate voltage $V_{GS}$ at room temperature. The on-state drain-source current can be as high as 35 µA µm$^{-1}$ when the FET operates at $V_{DS}$=1V. (**d**) $I_{DS}$ – $V_{DS}$ characteristics at room temperature. The source-drain current $I_{DS}$ varies linearly with $V_{DS}$ at sufficiently high gate voltages, indicating an Ohmic contact in the on states. (**e**) Drain-source current (on a logarithmic scale) as a function of top gate voltage with drain-source voltage of 0.1V, 0.5V and 1V. For a drain-source bias voltage $V_{DS}$ = 100 mV, an on-current of 4 µA µm$^{-1}$ is observed. At drain-source bias voltage $V_{DS}$ = 1V, the maximum on-current is 35 µA µm$^{-1}$. The on/off current ratio $I_{ON}/I_{OFF}$ can exceed 10$^9$ with an off-state current of about 10$^{-8}$ µA µm$^{-1}$. The subthreshold swing is around 0.7 V per decade. (**f**) Drain–source current $I_{DS}$ as a function

of drain source bias voltage $V_{DS}$ at different top gate voltages. A well-defined saturation region is observed for all applied gate voltages.

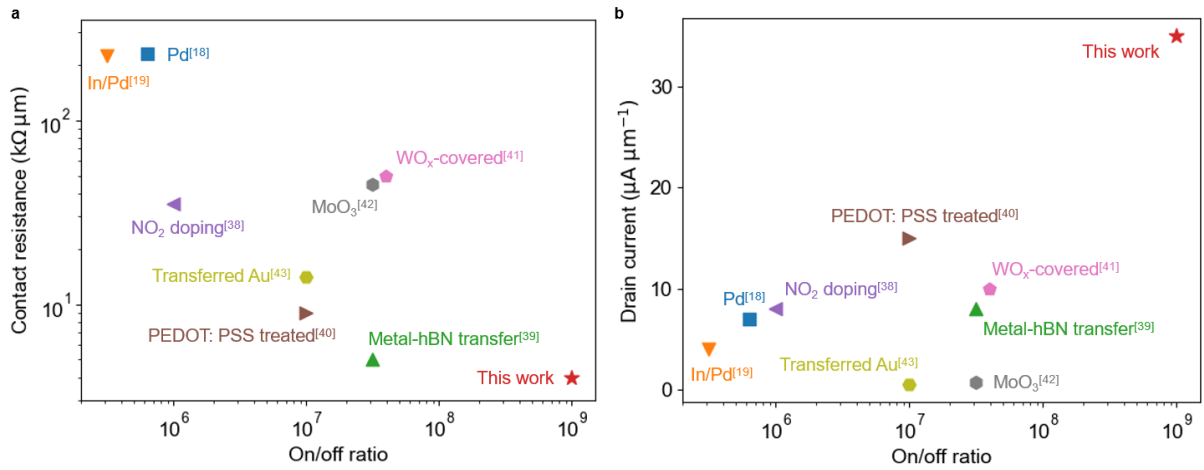

**Figure 5: Benchmark of p-type monolayer WSe$_2$ contact resistance and FET performance at room temperature and $|V_{DS}| = 1V$.** Comparison of contact resistance (a) and drain current (b) with values reported in the literature using different methods[18, 19, 38-43].

1. Chhowalla M, Jena D, Zhang H. Two-dimensional semiconductors for transistors. Nature Reviews Materials. 2016;1(11):1-15.
2. Novoselov KS, Mishchenko A, Carvalho oA, Castro Neto A. 2D materials and van der Waals heterostructures. Science. 2016;353(6298):aac9439.
3. Allain A, Kang J, Banerjee K, Kis A. Electrical contacts to two-dimensional semiconductors. Nat Mater. 2015;14(12):1195-205.
4. Kang S, Lee D, Kim J, Capasso A, Kang HS, Park J-W, et al. 2D semiconducting materials for electronic and optoelectronic applications: potential and challenge. 2D Materials. 2020;7(2).
5. Lin Z, Liu Y, Halim U, Ding M, Liu Y, Wang Y, et al. Solution-processable 2D semiconductors for high-performance large-area electronics. Nature. 2018;562(7726):254-8.
6. Li M-Y, Su S-K, Wong H-SP, Li L-J. How 2D semiconductors could extend Moore's law. Nature. 2019;567(7747):169-70.
7. Cao W, Kang J, Sarkar D, Liu W, Banerjee K. 2D Semiconductor FETs—Projections and Design for Sub-10 nm VLSI. IEEE Transactions on Electron Devices. 2015;62(11):3459-69.
8. Liu C, Chen H, Wang S, Liu Q, Jiang Y-G, Zhang DW, et al. Two-dimensional materials for next-generation computing technologies. Nature Nanotechnology. 2020;15(7):545-57.
9. Ma L, Nguyen PX, Wang Z, Zeng Y, Watanabe K, Taniguchi T, et al. Strongly correlated excitonic insulator in atomic double layers. Nature. 2021;598(7882):585-9.
10. Nguyen PX, Ma L, Chaturvedi R, Watanabe K, Taniguchi T, Shan J, et al. Perfect Coulomb drag in a dipolar excitonic insulator. arXiv preprint arXiv:230914940. 2023.
11. Qi R, Joe AY, Zhang Z, Xie J, Feng Q, Lu Z, et al. Perfect Coulomb drag and exciton transport in an excitonic insulator. arXiv preprint arXiv:230915357. 2023.
12. Qi R, Joe AY, Zhang Z, Zeng Y, Zheng T, Feng Q, et al. Thermodynamic behavior of correlated electron-hole fluids in van der Waals heterostructures. arXiv preprint arXiv:230613265. 2023.
13. Razavieh A, Zeitzoff P, Nowak EJ. Challenges and limitations of CMOS scaling for FinFET and beyond architectures. IEEE Transactions on Nanotechnology. 2019;18:999-1004.
14. Tung RT. The physics and chemistry of the Schottky barrier height. Applied Physics Reviews. 2014;1(1).
15. Louie SG, Cohen ML. Electronic structure of a metal-semiconductor interface. Physical Review B. 1976;13(6):2461.
16. Kim C, Moon I, Lee D, Choi MS, Ahmed F, Nam S, et al. Fermi level pinning at electrical metal contacts of monolayer molybdenum dichalcogenides. ACS nano. 2017;11(2):1588-96.
17. Sotthewes K, Van Bremen R, Dollekamp E, Boulogne T, Nowakowski K, Kas D, et al. Universal Fermi-level pinning in transition-metal dichalcogenides. The Journal of Physical Chemistry C. 2019;123(9):5411-20.
18. Wang Y, Kim JC, Li Y, Ma KY, Hong S, Kim M, et al. P-type electrical contacts for 2D transition-metal dichalcogenides. Nature. 2022;610(7930):61-6.
19. Wang Y, Kim JC, Wu RJ, Martinez J, Song X, Yang J, et al. Van der Waals contacts between three-dimensional metals and two-dimensional semiconductors. Nature. 2019;568(7750):70-4.
20. Shen PC, Su C, Lin Y, Chou AS, Cheng CC, Park JH, et al. Ultralow contact resistance between semimetal and monolayer semiconductors. Nature. 2021;593(7858):211-7.
21. Wang Y, Balgley J, Gerber E, Gray M, Kumar N, Lu X, et al. Modulation Doping via a Two-Dimensional Atomic Crystalline Acceptor. Nano Lett. 2020;20(12):8446-52.
22. Rizzo DJ, Jessen BS, Sun Z, Ruta FL, Zhang J, Yan JQ, et al. Charge-Transfer Plasmon Polaritons at Graphene/alpha-RuCl(3) Interfaces. Nano Lett. 2020;20(12):8438-45.

23. Biswas S, Li Y, Winter SM, Knolle J, Valentí R. Electronic Properties of α− RuCl 3 in proximity to graphene. Physical Review Letters. 2019;123(23):237201.
24. Rizzo DJ, Shabani S, Jessen BS, Zhang J, McLeod AS, Rubio-Verdu C, et al. Nanometer-Scale Lateral p-n Junctions in Graphene/alpha-RuCl(3) Heterostructures. Nano Lett. 2022;22(5):1946-53.
25. Sternbach AJ, Vitalone RA, Shabani S, Zhang J, Darlington TP, Moore SL, et al. Quenched Excitons in WSe(2)/alpha-RuCl(3) Heterostructures Revealed by Multimessenger Nanoscopy. Nano Lett. 2023;23(11):5070-5.
26. Pack J, Guo Y, Liu Z, Jessen B, Liu S, Holtzman L, et al. Improved p-type Contact to WSe 2 Using a α-RuCl 3 Charge-Transfer Interface. Bulletin of the American Physical Society. 2023.
27. Liu Y, Duan X, Shin HJ, Park S, Huang Y, Duan X. Promises and prospects of two-dimensional transistors. Nature. 2021;591(7848):43-53.
28. Fallahazad B, Movva HC, Kim K, Larentis S, Taniguchi T, Watanabe K, et al. Shubnikov-de Haas Oscillations of High-Mobility Holes in Monolayer and Bilayer WSe_2: Landau Level Degeneracy, Effective Mass, and Negative Compressibility. Phys Rev Lett. 2016;116(8):086601.
29. Movva HCP, Fallahazad B, Kim K, Larentis S, Taniguchi T, Watanabe K, et al. Density-Dependent Quantum Hall States and Zeeman Splitting in Monolayer and Bilayer WSe_2. Phys Rev Lett. 2017;118(24):247701.
30. Movva HC, Rai A, Kang S, Kim K, Fallahazad B, Taniguchi T, et al. High-mobility holes in dual-gated WSe2 field-effect transistors. ACS nano. 2015;9(10):10402-10.
31. Mashhadi S, Weber D, Schoop LM, Schulz A, Lotsch BV, Burghard M, et al. Electrical Transport Signature of the Magnetic Fluctuation-Structure Relation in alpha-RuCl(3) Nanoflakes. Nano Lett. 2018;18(5):3203-8.
32. Hattori Y, Taniguchi T, Watanabe K, Nagashio K. Anisotropic dielectric breakdown strength of single crystal hexagonal boron nitride. ACS Applied Materials & Interfaces. 2016;8(41):27877-84.
33. Li H, Ying Z, Lyu B, Deng A, Wang L, Taniguchi T, et al. Electrode-Free Anodic Oxidation Nanolithography of Low-Dimensional Materials. Nano Lett. 2018;18(12):8011-5.
34. Salahuddin S, Datta S, editors. Can the subthreshold swing in a classical FET be lowered below 60 mV/decade? 2008 IEEE International Electron Devices Meeting; 2008: IEEE.
35. Robertson J. High dielectric constant oxides. The European Physical Journal Applied Physics. 2004;28(3):265-91.
36. Zhang C, Tu T, Wang J, Zhu Y, Tan C, Chen L, et al. Single-crystalline van der Waals layered dielectric with high dielectric constant. Nat Mater. 2023.
37. Schwierz F. Graphene transistors. Nat Nanotechnol. 2010;5(7):487-96.
38. Fang H, Chuang S, Chang TC, Takei K, Takahashi T, Javey A. High-performance single layered WSe2 p-FETs with chemically doped contacts. Nano letters. 2012;12(7):3788-92.
39. Liu Y, Liu S, Wang Z, Li B, Watanabe K, Taniguchi T, et al. Low-resistance metal contacts to encapsulated semiconductor monolayers with long transfer length. Nature Electronics. 2022;5(9):579-85.
40. Zhang X, Kang Z, Gao L, Liu B, Yu H, Liao Q, et al. Molecule-Upgraded van der Waals Contacts for Schottky-Barrier-Free Electronics. Adv Mater. 2021;33(45):e2104935.
41. Yamamoto M, Nakaharai S, Ueno K, Tsukagoshi K. Self-Limiting Oxides on WSe2 as Controlled Surface Acceptors and Low-Resistance Hole Contacts. Nano Lett. 2016;16(4):2720-7.
42. Chen Y-H, Xing K, Liu S, Holtzman LN, Creedon DL, McCallum JC, et al. P-Type Ohmic Contact to Monolayer WSe2 Field-Effect Transistors Using High-Electron Affinity Amorphous MoO3. ACS Applied

Electronic Materials. 2022;4(11):5379-86.

43. Kong L, Zhang X, Tao Q, Zhang M, Dang W, Li Z, et al. Doping-free complementary WSe2 circuit via van der Waals metal integration. Nature communications. 2020;11(1):1866.

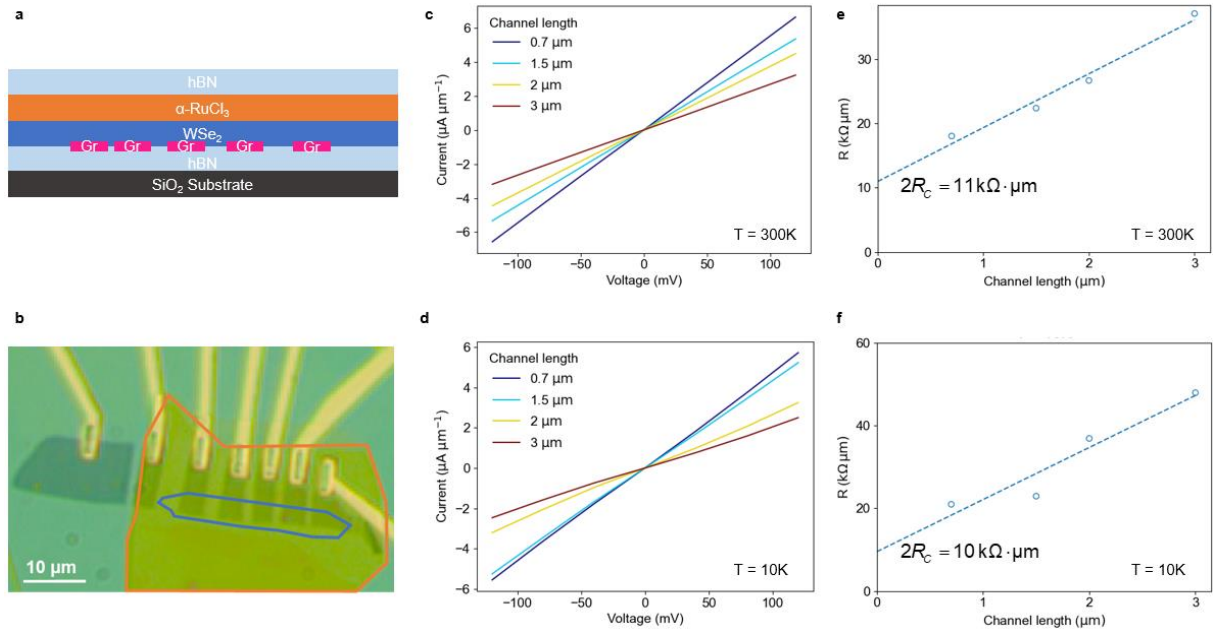

**Extended Figure 1: Contact resistance of Graphite/WSe$_2$/α-RuCl$_3$ contact.** (**a**) Schematic side-view of the device for TLM with graphite contacting the WSe$_2$ layer. (**b**) Optical image of the TLM device. The separation between electrodes ranges from 0.7 µm to 3 µm. The blue and orange shapes mark the monolayer WSe$_2$ and α-RuCl$_3$ region, respectively. (**c** and **d**) I-V curve for different channel length at room temperature (**c**) and T = 10K (**d**). Linear behavior is observed at both room temperature and low temperature, indicating the absence of a contact barrier. (**e**) and (**f**) Contact resistance R$_C$ extracted using TLM at room temperature (**e**) and T = 10K (**f**).

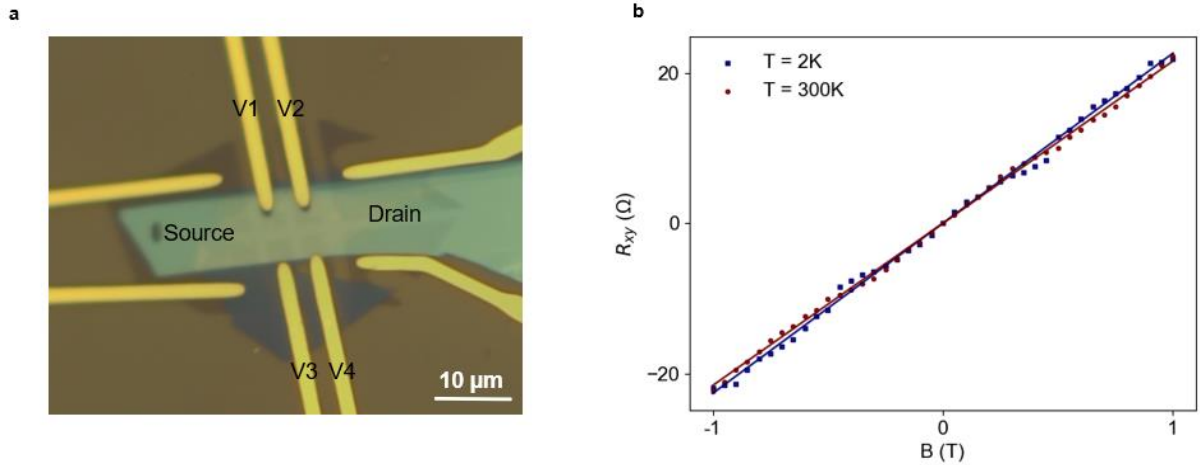

**Extended Figure 2: Determination of carrier density in WSe$_2$ contact region with a second monolayer WSe$_2$/α-RuCl$_3$ Hall bar device.** (**a**) Optical image of the multi-terminal monolayer WSe$_2$/α-RuCl$_3$ device for the Hall measurements. (**b**) Hall resistance $R_{xy}$ as a function of magnetic field where solid lines are linear fits to the experiment data. The positive sign of Hall slope confirms the hole doping in the monolayer WSe$_2$. We estimate a hole carrier density of $2.9\times10^{13}$ cm$^{-2}$ at both room temperature and low temperature from the Hall slope in this device.

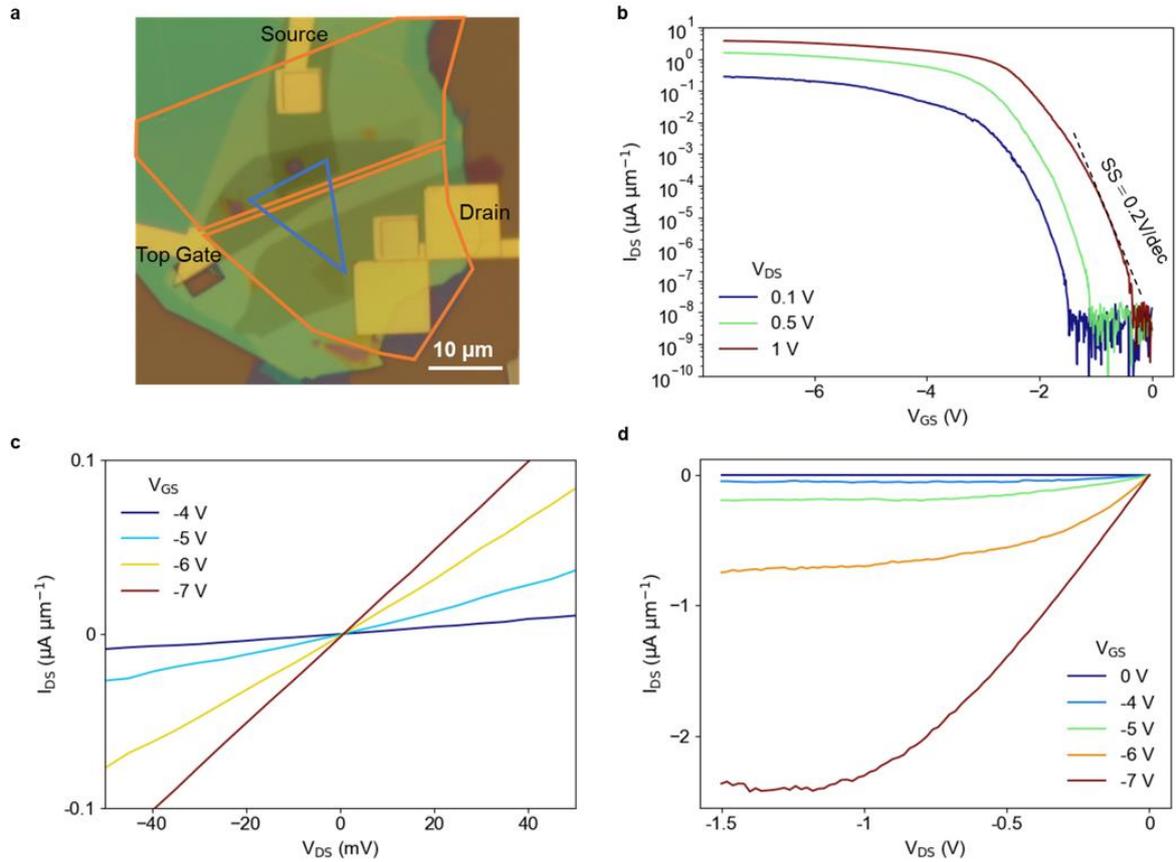

**Extended Figure 3: Characterization of a second monolayer WSe$_2$ FET device.** (**a**) Optical image of the monolayer WSe$_2$ FET device. Blue and orange boxes mark the monolayer WSe$_2$ region and α-RuCl$_3$ contact region, respectively. The channel length and width are about 0.5 μm and 8 μm, respectively. The top gate dielectric is a 15-nm hBN. (**b**) Drain-source current (on a logarithmic scale) as a function of top gate voltage with drain-source voltage of 0.1V, 0.5V and 1V. For a drain-source bias voltage V$_{DS}$ = 100 mV, an on-current of 0.3 μA μm$^{-1}$ is observed. The drain current on-off ratio I$_{ON}$/I$_{OFF}$ is about 10$^7$ with an off-state current of about 10$^{-8}$ μA μm$^{-1}$. At drain-source bias voltage V$_{DS}$ = 1V, the maximum on-current is 4 μA μm$^{-1}$ with an on/off current ratio I$_{ON}$/I$_{OFF}$ exceeding 10$^8$. The subthreshold swing is around 0.2 V per decade. (**c**) I$_{DS}$ – V$_{DS}$ characteristics at room temperature and the source-drain current I$_{DS}$ varies linearly with V$_{DS}$ in the hole side gate doping across all the gate voltages, indicating an Ohmic contact and a large degree of current control in our device. (**f**) Drain–source current $I_{DS}$ as a function of drain source bias voltage $V_{DS}$ at different top gate voltages. A well-defined saturation region is observed for all applied gate voltages.